\documentclass[prl,superscriptaddress,longbibliography,twocolumn]{revtex4}
\usepackage{graphicx}
\usepackage{bm}
\usepackage{color}
\newcommand{\be}{\begin{equation}}
\newcommand{\ee}{\end{equation}}

\newcommand{\br}{{{\bf{r}}}}

\newcommand{\bea}{\begin{eqnarray}}
\newcommand{\eea}{\end{eqnarray}}
\newcommand{\beal}{\begin{align}}
\newcommand{\eeal}{\end{align}}

\newcommand{\dg}{{\dagger}}
\newcommand{\pdg}{{\phantom\dagger}}

\begin{document}

\title{Hybridization-switching induced Mott transition in ABO$_3$ perovskites}
\author{Atanu Paul}
\affiliation{Department of Solid State Physics, Indian Association for the Cultivation of Science, Kolkata 700 032, India.}
\author{Anamitra Mukherjee}
\affiliation{School of Physical Sciences, National Institute of Science 
Education and Research, HBNI, Jatni 752050, India.}
\author{Indra Dasgupta}
\affiliation{Department of Solid State Physics, Indian Association for the Cultivation of Science, Kolkata 700 032, India.}
\author{Arun Paramekanti}
\affiliation{Department of Physics, University of Toronto, Toronto, Ontario, Canada M5S 1A7.}
\author{Tanusri Saha-Dasgupta}
\email{t.sahadasgupta@gmail.com}
\affiliation{Department of Condensed Matter Physics and Materials Science, S.N. Bose National Centre for Basic Sciences, Kolkata 700098, India.}
\affiliation{Center for Mathematical, Computational and Data Science, Indian Association for the Cultivation of Science, Kolkata 700 032, India.}

\date{\today}

\begin{abstract}
We propose the concept of ``hybridization-switching induced Mott transition'' which is relevant to a broad class of ABO$_3$
perovskite materials including BiNiO$_3$ and PbCrO$_3$ which feature extended $6s$ orbitals on the A-site
cation (Bi or Pb), and A-O covalency induced ligand holes. Using {\it ab initio} electronic structure
and slave rotor theory calculations, we show that such systems exhibit a breathing phonon driven A-site to oxygen hybridization-wave instability which conspires with strong correlations on the B-site transition metal ion (Ni or Cr) to induce a
Mott insulator. These Mott insulators with active A-site orbitals
are shown to undergo a pressure induced insulator to metal transition accompanied by a colossal volume collapse due
to ligand hybridization switching.
\end{abstract}

\maketitle

Recent advances in transition metal oxides have
led to a renewed interest in correlation driven metal-insulator transitions (MITs) \cite{mit-1,mit-2,mit-3,mit-4,mit-5,mit-6,vo-tio,cuo-nicl,zsa}.
Such MITs have been extensively explored in the ABO$_3$ perovskite family of materials, in which the
A-site and B-site cations live on interpenetrating (nominally) cubic lattices. In typical perovskites, the A-site ion is passive, while the B-site transition metal (TM) ion actively dictates
the electronic response. The A-site ion thus controls
electronic properties only {\it indirectly}: its size tunes B-O-B bond angles and thus the B-electron bandwidth, 
while its charge determines the
electron filling. For instance,
experimental
\cite{nickelate1,nickelate2,Ni3,Ni4,Ni5,Ni6,Chakhalian2013,Chakhalian2014,Sawatzky2014,Claessen2015,Ryan2015,Stemmer2015,
Triscone2016,Benckiser2016,Fabbris2016,vanderMarel2017,Review}
and theoretical \cite{Mazin2007,Balents1,Balents2,millis,anamitra,Millis2014,Millis2015,Georges2015,Sawatzky2016,Georges2017,Bibes2017}
studies of perovskite nickelates RNiO$_3$ (R being a rare-earth ion) have shown that
varying R, with increasing ionic sizes, induces a Mott insulator to metal transition driven by an increase in the Ni bandwidth.

In significant contrast to the above scenario, recent experiments on BiNiO$_3$ \cite{bno} and PbCrO$_3$ \cite{pco},
reveal a distinct behavior. BiNiO$_3$, with the large Bi cation at the A-site, 
is an insulator rather than a metal at ambient pressure\cite{footnote-bi}. This
insulator becomes metallic at a critical pressure of $3.5$ GPa, with a significant $2.5\%$ volume contraction. Similarly PbCrO$_3$ exhibits an insulator-to-metal transition at $2.5$ GPa \cite{pco} with a {\it colossal} $7.8\%$ volume collapse! Similar large volume shrinking insulator-to-metal transitions have been
observed upon heating; this technologically important phenomenon was termed ``colossal negative thermal expansion''  \cite{ncomm,takano}.

For BiNiO$_3$, attempts to address its MIT using Hartree-Fock theory \cite{motome} and 
dynamical mean-field theory \cite{dmft} have focused on models involving only Bi and Ni sites.
These studies view the insulator as a checkerboard
charge crystal [Bi$^{3+}_{0.5}$Bi$^{5+}_{0.5}$][Ni$^{2+}]$, assuming that Bi acts as a valence skipping ion with
an attractive $U$ Hubbard interaction, while the high pressure metal results from a valence transition
into a uniform [Bi$^{3+}$][Ni$^{3+}]$ configuration. However, photoemission spectroscopy 
on the metal \cite{pes,xas} reveals that the nickel valence state is far from being purely Ni$^{3+}$.
At the same time, Bi$^{5+}$ has an energetically deep $6s$ shell \cite{madelung} which
strongly suppresses Bi$^{3+}$-Bi$^{5+}$ charge disproportionation. 
These theories neither discuss the crucial role of ligand nor that of the lattice in the MIT.
These issues are reminiscent of the ``charge disproportionation'' debate in insulating RNiO$_3$ and B-site bismuthate BaBiO$_3$ \cite{haule,Ghosez,BaBiO_1,BaBiO_2,BaBiO_3,BaBiO_4}.
Indeed, rather than being Ni$^{2+}$-Ni$^{3+}$ charge crystals, the RNiO$_3$ insulators feature a NiO$_6$ breathing mode
instability, leading to
site-selective Mott insulators; holes on one Ni sublattice undergo a Mott transition
while holes on the other sublattice reside in NiO$_6$ molecular orbitals \cite{anamitra,millis,haule,Ghosez}.


In this Letter, we show that a natural solution to volume collapse MIT in TM ABO$_3$ with active A site 
emerges if the oxygen sites and lattice distortions are
explicitly included in 
modelling these novel perovskites.
In steps towards this, we first show, using density functional theory 
(DFT) on specific ABO$_3$ perovskites, BiNiO$_3$ and PbCrO$_3$, that the key to their phenomenology
lies in an active A-site with $6s$ orbital which can strongly hybridize with oxygen,
  creating ligand holes. Based on this, we propose a simplified 
model for such perovskites which includes all three ions (A,B,O) and the lattice degree of freedom, and 
solve it using slave rotor theory. We find that as the oxygen energy level becomes close
to that of A-site cation, driving a negative charge-transfer situation, the extended A-site orbital permits an A-O breathing mode
instability, which leads to a three-dimensional (3D) checkerboard pattern of compressed and expanded 
AO$_{12}$ polyhedra. This is analogous to a 3D Peierls' transition in a strongly correlated regime, which
does not rely on Fermi surface nesting \cite{pr-1,pr-2}.
This instability suppresses the B-O hybridization, triggering a Mott insulating state due to 
strong correlations on the B-site ions.
Applying pressure shifts the ligand energy, favoring 
B-O over A-O hybridization. This {\it hybridization-switch} eventually suppresses the A-O breathing mode instability and the
concomitant hybridization-wave, leading to a metal 
dominated by B-O states at the Fermi level.
Our results motivate us to conclude that BiNiO$_3$ and PbCrO$_3$
belong to a broad category of compounds exhibiting {\it hybridization-switching induced 
Mott transition.} We propose further material
candidates in this category, namely TlMnO$_3$ and InMnO$_3$.
 
{\it Pressure-induced structural transition.---} At ambient pressure (AP), the crystal structure of BiNiO$_3$ is triclinic, with two inequivalent Bi sites (Bi1, Bi2),
and four inequivalent Ni sites. The 
high pressure (HP) phase, above $3.5$GPa \cite{jacs}, has orthorhombic symmetry, with equivalent Bi and Ni sites. 
The AP phase has a staggered pattern of compressed and expanded BiO$_{12}$ polyhedra, while the 
HP phase features BiO$_{12}$ polyhedra of uniform volume. 
We begin by describing this structural transition within DFT.
DFT calculations were carried out in pseudo-potential plane-wave basis with generalized gradient approximation \cite{pbe} 
with Hubbard $U$ (GGA+$U$) \cite{gga+u} ($U \!\!=\!\! 4$ eV, $J_H \!\!=\!\! 0.9$eV), as implemented in Vienna-Abinitio-Simulation-Package \cite{vasp};
see Supplemental Material (SM) for details \cite{suppl}.

We fit the volume dependence of our DFT cohesive energies of the AP (triclinic)
and HP (orthorhombic) structures to the Birch-Murnaghan equation \cite{b-m}. 
Using a common tangent construction (see Fig.~1 (a)), we find
a transition from low-symmetry triclinic to high-symmetry orthorhombic structure
with a volume reduction of $\approx$ 3$\%$, in good agreement with high pressure 
experiments \cite{ncomm}.

To understand the role of the bond deformation in the MIT, we 
computed the stiffness of the Bi-O bonds. Starting from the
undistorted orthorhombic structure at volumes corresponding to AP and HP (at 6 GPa), and
replacing Ni ions by a uniform
positive background, we calculated the energy change $\delta E$ for
small breathing displacement of oxygen atoms ($\delta$O) from their equilibrium positions; see Fig.~1(b) inset.
Fitting $\delta E \!=\! \frac{1}{2} k (\delta {\rm O})^2$ yields the Bi-O bond stiffness constants 
$k_{\rm BiO} \!=\! 2.06 {\rm eV}/\AA^{2}$ and $2.32 {\rm eV}/\AA^{2}$ for AP and 6 GPa HP volumes, respectively.
For the Ni-O sublattice, see Fig.~1(b), corresponding calculations yield $k_{\rm NiO} \!=\! 7.96 {\rm eV}/\AA^{2}$ and $10.84 {\rm eV}/\AA^{2}$. 
Thus, the Ni-O bond, which is stiffer than Bi-O bond, becomes substantially stiffer at HP, suppressing a
  distortion of the Ni-O sublattice. For comparison, the similarly calculated
  Ni-O stiffness in PrNiO$_3$, which shows a breathing mode distortion \cite{prnio},  is $ 7.22{\rm eV}/\AA^{2}$, a factor of $1.5$ 
  smaller than that of HP BiNiO$_3$. This explains the
absence of a breathing mode distortion of NiO$_6$ octahedra in the HP phase, and the resulting stability of the volume-collapsed metal against a Ni ``charge-disproportionation'' MIT.

\begin{figure}
\includegraphics[width=0.45\textwidth,keepaspectratio]{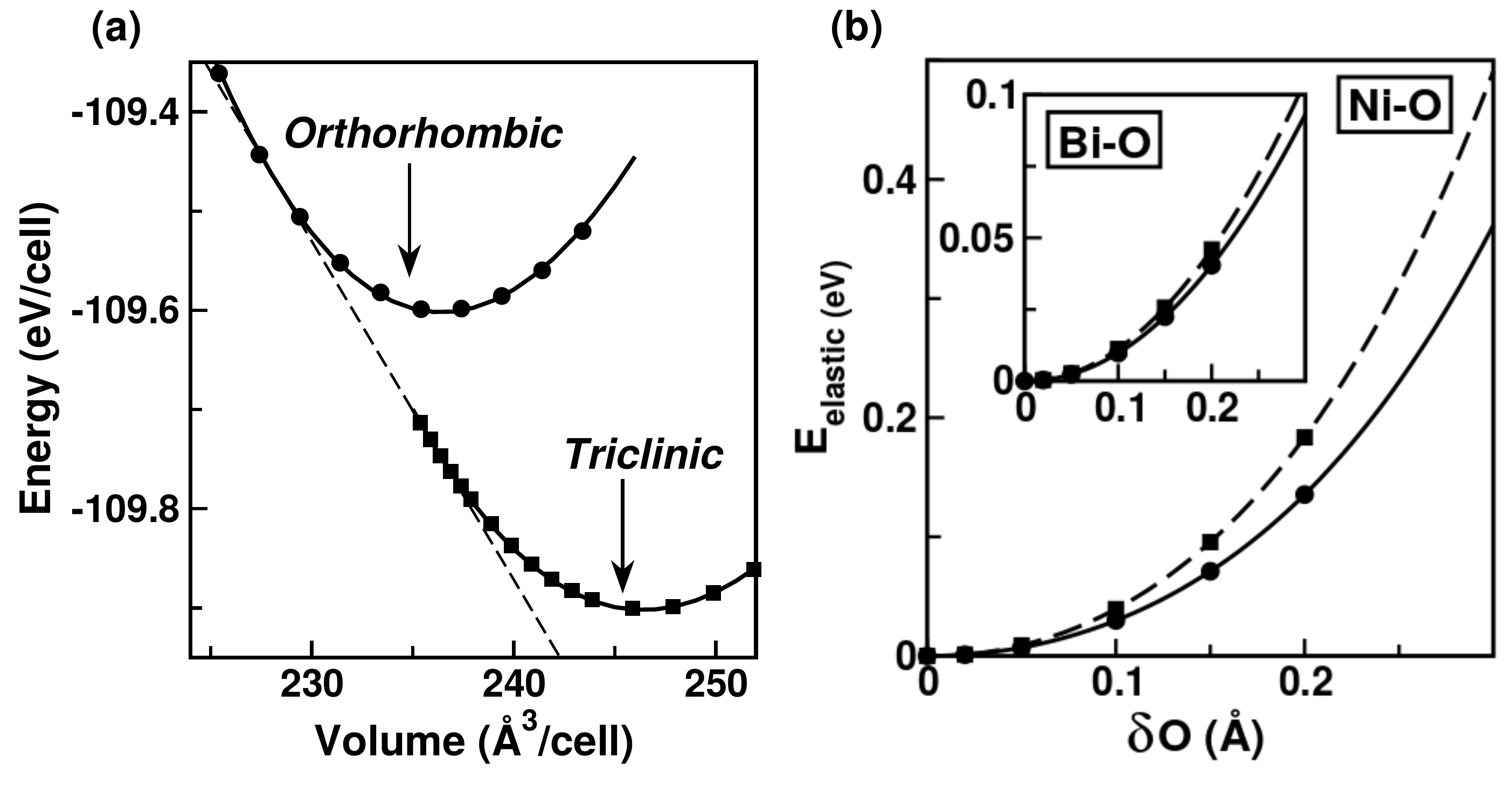}
\caption{(Color online) (a) DFT cohesive energy versus volume for AP (triclinic) and HP (orthorhombic) 
BiNiO$_3$; the intersection points of the plotted common tangent with the two curves yields the volume change $\Delta V/V \approx 3\%$ at the transition.
 (b) Elastic energy of Ni-O sublattice as a function of O-displacement for the HP (dashed) and AP (solid) volumes.
Inset shows similar plot for Bi-O sublattice.}
\end{figure}  

{\it DFT electronic structure.---} Fig.~2(a) shows the spin-polarized GGA+$U$ density of states (DOS) of AP and HP ($7$GPa) BiNiO$_3$, respectively,
projected onto Bi-$s$, Ni-$d$ and O-$p$ states.
The Ni-$d$ and O-$p$ plots are the DOS averaged over four inequivalent Ni sites in AP, and six and two O sites in AP and HP, respectively.

At AP, our DFT calculation gives rise to an insulating solution upon including
antiferromagnetic (AFM) order.
Both Bi1 and Bi2 sites show filled $6s$ states deep down 
in energy $\approx\! 10.5$eV below Fermi level ($E_F$). The split-off, unoccupied part of Bi2-$s$ states, 
which is at $\approx\! 1$eV above $E_F$, and well-separated from the 
filled Bi2-$s$ states by $\approx 9$-$10$eV, is entirely derived from strong admixture with 
O-$p$ states, signaling creation of ligand holes (see encircled regions in the figure). Thus, the insulator is {\it weakly} Bi-charge disproportionated \cite{foot-cd}.
The contribution of Ni-$d$ to this split-off state is small.
We find the Ni-$d$ states are filled in the majority spin channel, while in the minority spin channel
the octahedral crystal-field split Ni-$t_{2g}$ and Ni-$e_g$ states are
respectively filled and empty (positioned beyond the energy shown in the figure).
This suggests the stabilization of [Bi1$^{3+}_{0.5}$(Bi2$^{3+}\underline{L}^{2(1-\delta)}$)$_{0.5}$][Ni$^{2+}\underline{L}^{\delta}$] configuration in AP, instead of the
proposed [Bi1$^{3+}_{0.5}$Bi2$^{5+}_{0.5}$][Ni$^{2+}$] configuration \cite{motome,dmft}.
Our calculated oxygen magnetic moment is large $\approx \! 0.1 \mu_B$, 
contrary to the expectation of non-magnetic O$^{2-}$ for [Bi1$^{3+}_{0.5}$Bi2$^{5+}_{0.5}$][Ni$^{2+}$] configuration.
Total energy calculations show that Ni favors a G-type AFM order, in agreement with neutron diffraction \cite{jacs}.

At HP, the DFT calculation gives rise to a metallic solution, with dispersive bands 
crossing the Fermi level.
Our DFT total energy calculations shows ferromagnetic Ni-Ni interactions; we therefore
predict the HP metal should show ferromagnetic correlations. Analyzing the projected O-$p$ DOS, we again find significant weight at the
unoccupied part, reflecting the ligand hole. However, the unoccupied O-$p$ states
have lot more Ni-$d$ character, and much less Bi-$s$ character, compared to the AP phase. Interestingly, the magnetic
moment on Ni site in the HP phase ($1.48 \mu_B$) is not much smaller than in
the AP phase ($1.73 \mu_B$), in marked contrast to proposal of Ni$^{2+}$ to Ni$^{3+}$ valence
transition between AP and HP. DFT thus suggests stabilization of [Bi$^{3+}\underline{L}^{\delta}$][Ni$^{2+}\underline{L}^{1-\delta}$]
configuration in the metal. This picture within single-reference description of DFT is close to the multi-reference description given by Ni K-edge X-ray absorption spectroscopy.\cite{xas}
Calculation of crystal orbital Hamiltonian population (COHP) \cite{cohp1,cohp2} (see SM), corroborates the change of ligand hole character from
Bi-$s$ to Ni-$d$.

\begin{figure}
\includegraphics[width=8.7 cm,keepaspectratio]{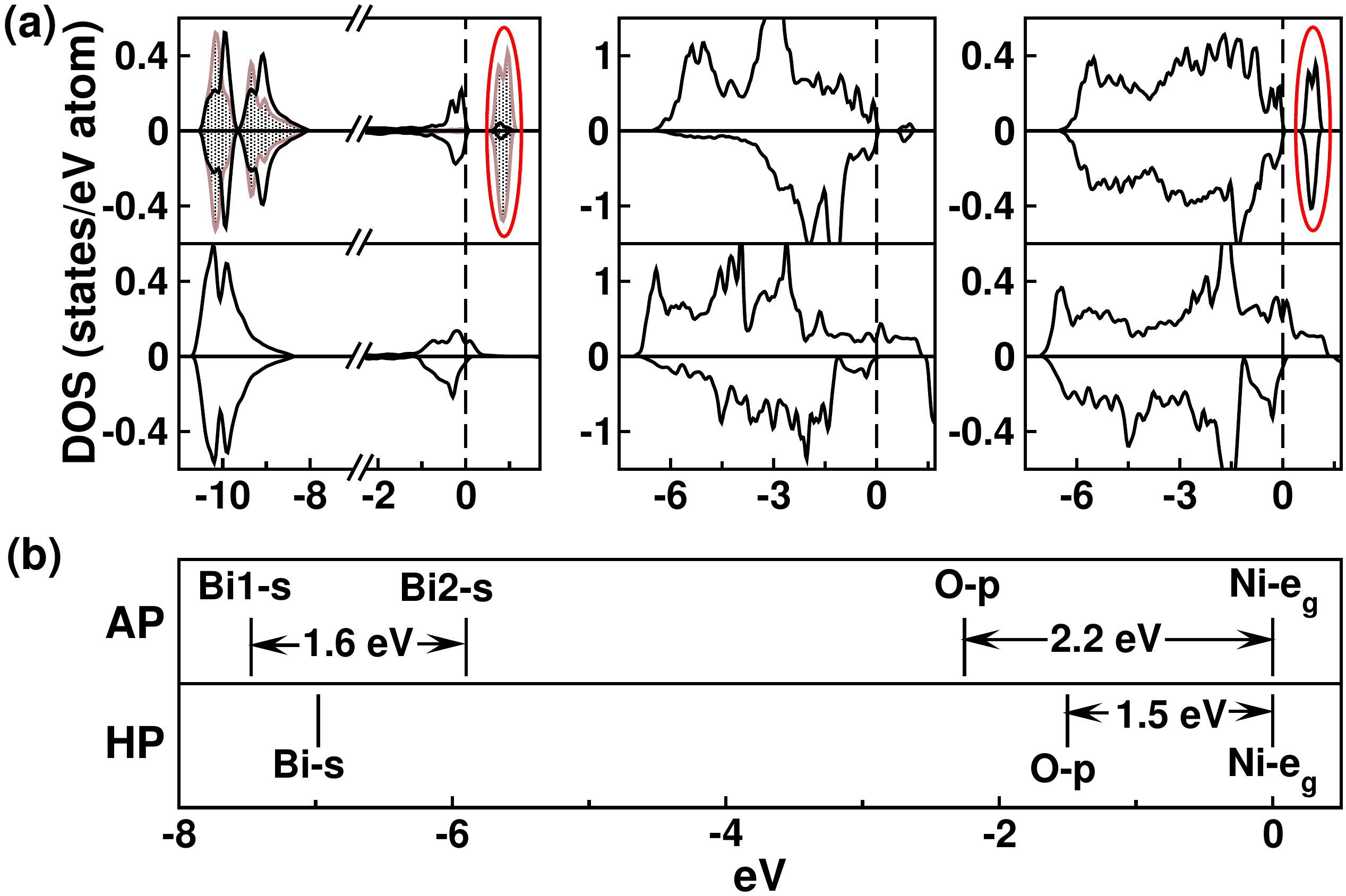}
\caption{(Color online) (a) GGA+$U$ projected DOS of BiNiO$_3$ in AP (top panels) and HP (bottom panels)
  phase. Left, middle, and right panels show projections to Bi-$s$ (displaying relevant energy ranges), Ni-$d$, and O-$p$. Zero of energy is set at GGA+$U$ Fermi energies. For AP, projections to Bi1 (solid, black) and Bi2 (shaded) are
  shown separately. (b) Calculated energy levels of Bi-$s$, O-$p$ and Ni-$e_g$ in AP and HP phase. 
Zero of energy is set at Ni-$e_g$.}
\end{figure} 


What causes this shift of oxygen covalency? To answer this, we show in Fig. 2(b), the computed
Bi-$s$, Ni-$e_g$ and O-$p$ energy level positions in AP and HP phases, obtained from
the low-energy tight-binding Hamiltonian in Wannier function basis within the N-th order muffin-tin-orbital (NMTO) 
formulation of downfolding technique \cite{nmto,footnote} (see SM).
At AP, the $s$-level energy positions of Bi1 and Bi2, differ by about $1.5$eV, Bi2-$s$ being closer
to O-$p$ compared to Bi1-$s$, leading to stronger covalency between Bi2-$s$ and O-$p$. Ligand holes are thus preferably associated
with Bi2. At HP,
the energies of
Ni-$d$ and O-$p$ get markedly closer, driving a covalency shift to Ni-O.

Our DFT results for PbCrO$_3$ are qualitatively similar. The quantitative
differences in PbCrO$_3$, a smaller critical pressure and larger volume collapse, arise from the O level
lying closer to Pb. Thus, the Pb-O covalency is stronger than Bi-O. At the same time,
Cr $t_{2g}$ orbitals hybridize less effectively with O than Ni $e_g$  (see SM).

{\it Slave rotor theory. ---} In order to go beyond the DFT+$U$ treatment of strong correlation effect, and
  capture the Mott transition without any assumption of magnetic ordering, we next investigate such ABO$_3$ perovskites in a DFT-inspired ``$s$-$p$-$d$'' model,
which we study using slave rotor mean field theory \cite{Florens2002,Florens2004,Lau2013, Lee2005, Zhao2007}.
Our work 
represents a novel application of slave-rotor
which simultaneously treats all three ions (A, B, O).
In contrast to previous work \cite{motome,dmft}, 
our model does not include a phenomenological {\it negative} $U$ on
the A-site. Instead, we
include phonon distortion and A-O hybridization, which provides a more meaningful microscopic picture
\cite{negu,harrison}.

\begin{figure}
\includegraphics[width=8.7cm,keepaspectratio]{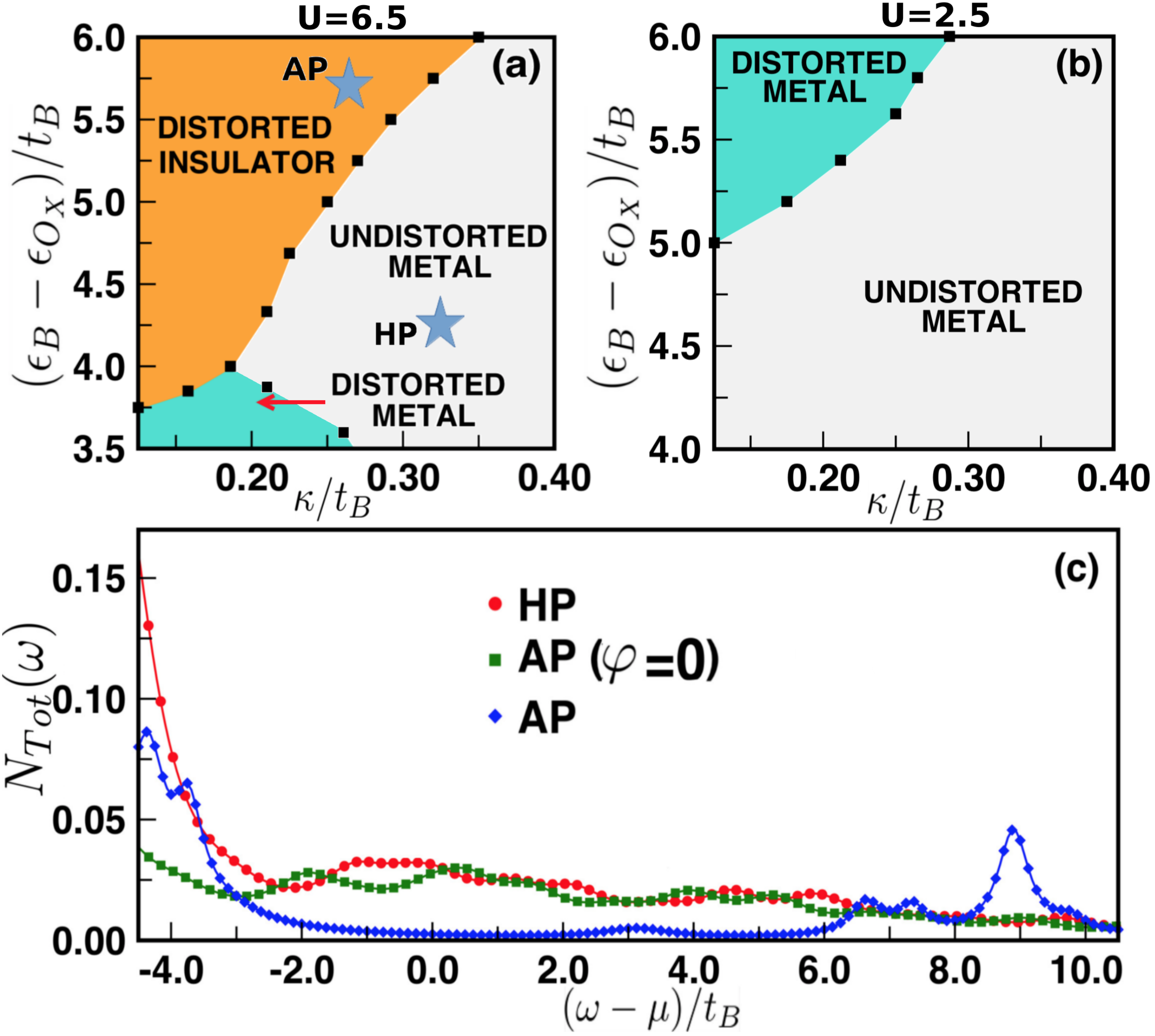}
\caption{(Color online) Phase diagram varying $\kappa$ and $\epsilon_B-\epsilon_{\rm Ox}$ for (a) $U/t_B\!=\! 6.5$
and (b) $U/t_B\!=\! 2.5$.
(c) DOS $N_{\rm Tot}(\omega)$ for typical points in AP and HP phases, marked by stars in (a),
as well as DOS for metastable AP phase with imposed $\varphi=0$.}
\end{figure}

Our model for ABO$_3$
consists of a multi-orbital manifold on the B-site, with non-degenerate orbitals on A and on the oxygen site.
The on-site energies are denoted by $\epsilon_{\rm A}$, $\epsilon_{\rm B}$, and $\epsilon_{\rm Ox}$;
we fix $\epsilon_{\rm B}=0$. Denoting A-O and B-O hopping amplitudes in the symmetric phase 
as $t_{\rm A}$, $t_{\rm B}$ respectively, and including a Hubbard $U>0$ on B-site, yields
the Hamiltonian $H \!=\! H_1 + H_2 + H_3 + H_4$, with
\begin{eqnarray*}
H_1 &=&
\epsilon_{\rm A} \sum_{\br,\sigma} a^\dg_{\br+\Delta,\sigma} a^\pdg_{\br+\Delta,\sigma}+
\epsilon_{\rm Ox} \sum_{\br,\sigma,\delta} \ell^\dg_{\br+\delta,\sigma} \ell^\pdg_{\br+\delta,\sigma} \\
H_2 &=& - t_{\rm B} \sum_{\br\alpha\sigma\delta} g_{\alpha\delta} 
(b^\dg_{\alpha,\br,\sigma} [\ell^\pdg_{\br+\delta,\sigma} + \ell^\pdg_{\br-\delta,\sigma}] + {\rm h.c.}) \nonumber \\
&+& \frac{U}{2} \sum_{\br} (\sum_{\alpha\sigma} b^\dg_{\alpha,\br,\sigma} b^\pdg_{\alpha,\br,\sigma}-2)^2 \\
H_3 &=& - t_{\rm A} \!\!\!\!\! \sum_{\br,\delta,\eta_\delta,\sigma} \!\! [1 + \varphi (-1)^\br] (a^\dg_{\br+\Delta,\sigma} \ell^\pdg_{\br+\Delta+\eta_\delta,\sigma} \!+\! {\rm h.c.})\\
H_4 &=& 12 N \times \frac{1}{2} \kappa \varphi^2
\end{eqnarray*}
where $a$, $b$, $\ell$ denote electron operators on A, B, and ligand site respectively, with $\alpha$ labelling B-site orbitals. 
Here $H_1$ describes the on-site energy, with
a choice $\epsilon_B\!=\!0$, while $H_2$ and $H_3$ respectively 
describe the B-O and A-O electronic Hamiltonians, and
$H_4$ denotes the elastic energy cost of A-O bond deformations. Based on DFT,
we assume the stiffer B-O sublattice to be immune to breathing distortion.
Staggered A-O hopping, $t_{\rm A} (1\pm \varphi)$, permits us to capture the A-O {\it hybridization-wave}. 
In the symmetric phase $\varphi \!=\! 0$. In the distorted phase 
$\varphi\! =\! \beta \frac{\delta a_{\rm AO}}{a_{\rm AO}}$, where $\beta \! \equiv\! (\partial \ln t_{\rm A}/\partial \ln a_{\rm AO})$,
and $\delta a_{\rm AO}$ is the change in A-O bond length compared to its undistorted value $a_{\rm AO}$.
The elastic energy cost in $H_4$ is $\frac{1}{2} \kappa \varphi^2$ per bond, where $\kappa \!=\! k a_{\rm AO}^2/\beta^2$,
with spring stiffness constant $k$,
and $12 N$ A-O bonds. For BiNiO$_3$, we have two $e_g$ orbitals $(1 \equiv d_{x^2-y^2}$, $2 \equiv d_{3z^2-r^2})$ at the 
Ni-site, with $g_{1,x/y/z} \!=\! \{1,-1,0\}$ and $g_{2,x/y/z} \!=\! \frac{1}{\sqrt{3}} \{-1,-1,2\}$ due to orbital-dependent Ni to ligand hopping \cite{sk}. 
This model can be extended to study $t_{2g}$ orbitals relevant to Cr in PbCrO$_3$.

\begin{figure}
\includegraphics[width=8.5 cm,keepaspectratio]{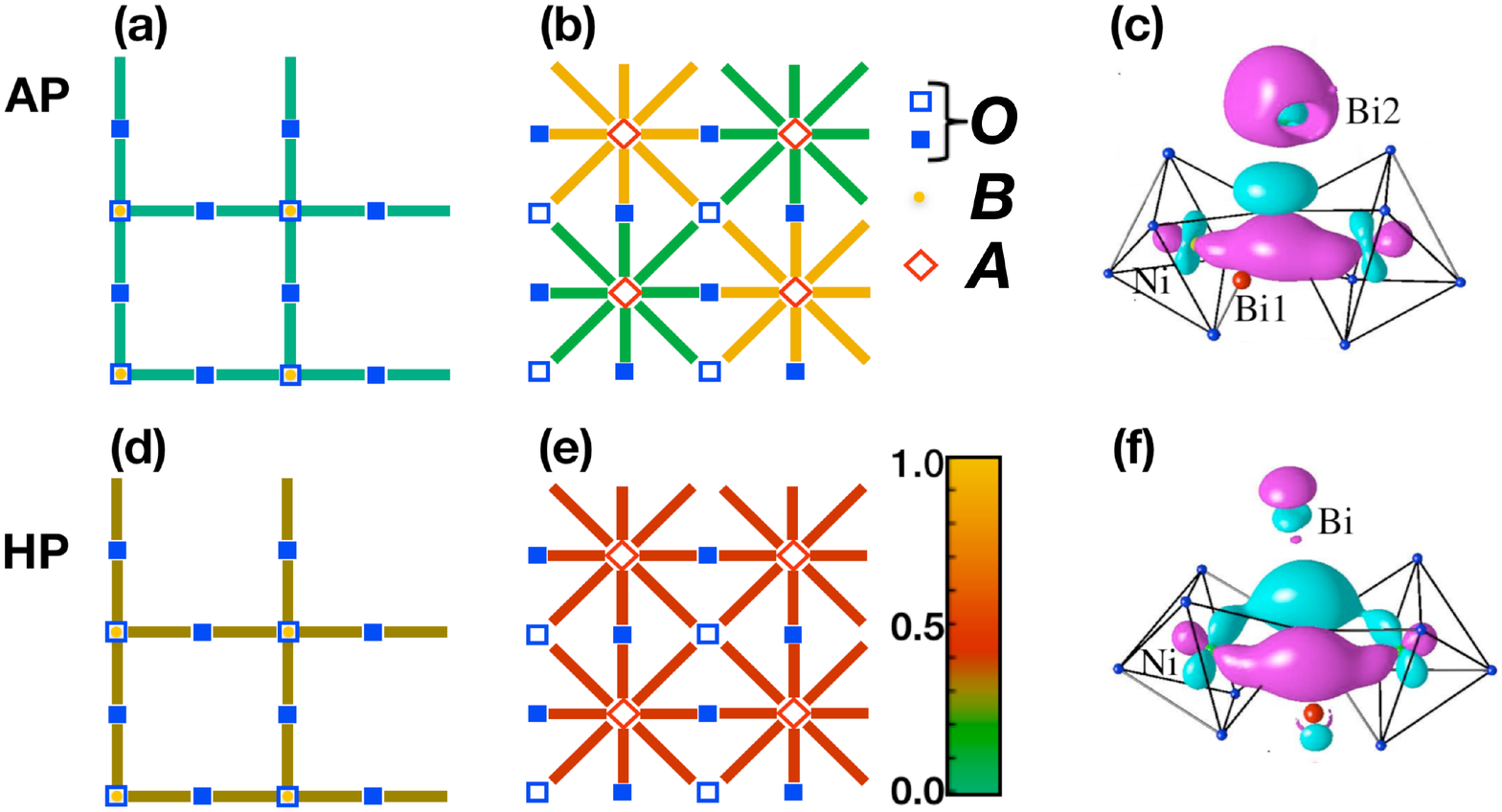}
\caption{(Color online) Bond-dependent kinetic energy from model (color bar shows magnitude) for
B-O and A-O sublattices in AP ((a) and (b)) and HP ((c) and (d)) phase, projected to $xy$ plane.
Constant amplitude surfaces of DFT O-$p$ Wannier functions for BiNiO$_3$ at AP (e) and HP (f), superposed on
NiO$_6$ octahedra with adjacent two Bi ions. Cyan (light) and magenta (dark) colors 
indicate opposite signs.}
\label{theory-pd}
\end{figure}

We study the zero temperature phase diagram of Hamiltonian $H$ using slave rotor mean field theory on the B-site (see SM for details). 
This approach ignores fluctuations of the rotor field as well as gauge fluctuations;
nevertheless, it 
reasonably captures
strong correlation effects \cite{Florens2002, Florens2004, Lau2013, Lee2005, Zhao2007}.
We work in units with $t_B\!=\!1$. 
We choose $(\epsilon_{\rm A}\!-\! \epsilon_{\rm B})/t_B\!=\! 8$, $t_A/t_B\!=\! 2.5$,
and vary $\kappa$ and $\epsilon_{\rm Ox}$, with $\epsilon_{\rm A} \!<\! \epsilon_{\rm Ox} \!<\! \epsilon_{\rm B}$.
Some variation in these parameters does not qualitatively affect our main results.
Fig.~3 shows the phase diagram with varying ligand energy and stiffness $\kappa$, for a B-site ion with 
(a) $U/t_B \!=\! 6.5$, and (b) $U/t_B \!=\! 2.5$.
For significant $\kappa$, when $\epsilon_{\rm Ox}$ is close to $\epsilon_{\rm B}$, the ground state is
an undistorted metal (UM) with non-integer B-site occupancy $n_{\rm B}$. Tuning $\epsilon_{\rm Ox}$ 
towards $\epsilon_A$, or $\kappa$ to smaller values, leads to a transition into either a
distorted Mott insulator (DI) with pinned $n_{\rm B}=2$ (i.e., a ``doping tuned'' Mott transition on B site), or a distorted metal (DM), depending on $U$.
The strongly correlated case, $U/t_B\!=\!6.5$, which results in a Mott localized DI, 
mimics the Ni site in BiNiO$_3$; for $t_B \!=\! 0.75$eV, we get $U \! \approx \! 5$eV.
A spontaneous A-O hybridization-wave $\pm \varphi$ thus cooperates with the large $U$ on 
the B-site, and results in this B-O to A-O \textit{hybridization-switching induced Mott insulator}.

The critical value $(\epsilon_B-\epsilon_{\rm Ox})_{\rm crit}$ for the MIT increases with increasing $\kappa$.
At the indicated point in the DI phase in Fig.~3(a), with $\kappa \! \approx \! 0.25$, the optimal distortion $\varphi \approx 0.55$. 
Choosing $\beta \!\approx\! 5$ \cite{Harrisonscaling} and $a_{\rm BiO}=2.3\AA$ yields $\delta a_{\rm AO} \! \approx \! 0.25\AA$ as seen in BiNiO$_3$. Setting $t_B=0.75{\rm eV}$,
we get $(\epsilon_B-\epsilon_{\rm Ox})_{\rm crit}\! \approx \! 3.8{\rm eV}$ for metallization, while the $\kappa$ value implies a
stiffness $k \! \approx \! 1 {\rm eV}/\AA^2$, in reasonable agreement with DFT given
the simplicity of the model which retains only Ni $e_g$ and a single ligand orbital.
The DOS, shown in Fig.~3 (c), displays a Mott gap in the DI phase while it is metallic in the UM phase.
Forcing $\varphi$ = 0 within the DI phase leads to a metallic DOS (see Fig.~3(c));
correlations alone are thus {\it insufficient} to induce an insulator. 

{\it Bond dependent hybridization. ---} Figs.~4 (a)-(d) shows the bond-dependent kinetic energy on the A-O
and B-O bonds from the model Hamiltonian.
In the DI phase, the B-O hybridization is weak, while
AO$_{12}$ polyhedra display the hybridization-wave. In the UM phase,
on the other hand, the B-O hybridization strengthens significantly compared to that in insulating phase, 
while the A-O hybridization becomes uniform. We corroborate this using
NMTO-downfolding-derived DFT Wannier function plots of BiNiO$_3$ in 
O-$p$ only basis calculations, as shown in Fig. 4(e), (f) \cite{foot2}.
At AP, the Wannier function is highly asymmetric, having a
pronounced tail at Bi2 and nearly vanishing at Bi1. At HP, on the other hand, it is symmetric
between Bi sites. 
Moving from AP to HP
the tail at Ni is strengthened significantly,
highlighting the change from Bi $s$-like to Ni $d$-like ligand hole.

{\it Conclusion. ---} 
We have proposed the concept of a hybridization-switching induced Mott transition
in ABO$_3$ perovskites, with
BiNiO$_3$ and PbCrO$_3$ as concrete examples.
Using DFT and slave rotor theory, we have identified its key ingredients as: (a)
extended A-site orbitals, which strongly hybridize with oxygen to generate ligand holes
and highly covalent A-O bonds susceptible to distortion, and (b) strong correlations on B-site ion.
Pressure tuning the oxygen energy produces a volume-collapse Mott insulator to metal transition 
via shift in covalency.
Based on our study, we propose TlMnO$_3$ \cite{tmo,footnote3}, and even $5s$ systems like InMnO$_3$, as
further promising material candidates \cite{footnote4}.
Charge doping such Mott
insulators may lead to polaronic transport and superconductivity.

TS-D thanks the Department of Science and Technology, India for the support through Thematic Unit 
of Excellence. AP is supported by NSERC of Canada and the Canadian Institute
for Advanced Research. ID is supported by Department of Science and Technology, India.


\begin{thebibliography}{10}
\bibitem{footnote-bi} Note in case of BiNiO$_3$, Bi ion occupies the A-site, providing
an intricate interplay between active A-site and the active strongly correlated B-site, namely Ni. 
This is unlike the case of BaBiO$_3$ (see e.g. Refs.\cite{BaBiO_1,BaBiO_2,BaBiO_3,BaBiO_4}) 
in which the Bi ion, occupying the B-site, is 
at the center of the oxygen octahedra. For BaBiO$_3$, the structure and phenomenology is thus more similar to the rare-earth 
nickelates, with Ba cation not playing any active role.
\bibitem{footnote} Ni-$t_{2g}$ being fully filled in both AP and HP phases, they have not been considered. The small, but finite splitting between the two e$_g$ levels of Ni, and three $p$ levels of oxygen due to finite distortion is not shown for simplicity. Also there is some variation between energy levels of different inequivalent Ni's (in case of AP)
and different inequivalent O's. Only the averaged positions are shown.
\bibitem{foot2}  The Wannier functions are constructed
by NMTO-downfolding technique, in which only O-$p$ degrees of freedom  are kept active in the basis, rest including
Ni-$d$ and Bi-$s$ degrees of freedom being downfolded.
\bibitem{footnote3} The ambient condition crystal structure of TlMnO$_3$ reported in literature \cite{tmo} has triclinic symmetry with
two inequivalent Tl positions. Although the valence of Tl was assumed to be nominally $3+$, the first-principles calculation reported in \cite{tmo}
shows that the energies of the formally `unoccupied' Tl 6s states are substantially lower than expected. Thus the Tl 6s states, which would be
completely empty in the ionic limit, develop significant occupation through Tl-O covalency and form the bottom of the valence band.
The situation is thus analogous to BiNiO$_3$, and a hybridization wave would result in a lower symmetry structure
with two inequivalent Tl sites. The case of TlMnO$_3$ should thus be revisited in light of our findings, to explore the effect of 
pressure, temperature, and isovalent substitution on Tl site.
\bibitem{footnote4} Note that discussed cases are different
in behavior from that of BiFeO$_3$ or PbTiO$_3$, the so-called magnetic ferroelectric compounds, though they also possess A site with extended 6s orbital. What makes BiFeO$_3$ or PbTiO$_3$ different from BNO or PCO, is the absence of ligand hole in the former and presence of ligand hole in the later. Both BNO and PCO are in negative-charge transfer situation at AP, thereby creation of ligand hole in A-O sublattice, and subsequent ligand-hole driven hybridization wave. Unlike this, though the finite A-O covalency in case of BiFeO$_3$ or PbTiO$_3$ drives the stereochemical activity of A sites, it is not sufficiently strong to drive the negative charge transfer situation.
\end{thebibliography}
\end{document}